\documentclass[onecolumn,amsmath,amssymb,nofootinbib,12pt]{article}
\pdfoutput=1 % if your are submitting a pdflatex (i.e. if ∞ you have
\usepackage{jheppub} % for details on the use of the package, please
\usepackage[multiple]{footmisc}

\usepackage{tensor}
\usepackage{graphicx,subcaption}
\graphicspath{ {figures/} }	
\usepackage{amsfonts}
\usepackage{xparse}
 \usepackage{amsmath}
 \usepackage{amssymb}
 \usepackage{mathtools}
\usepackage{epsfig}
\usepackage{pdfpages}
\usepackage{bbm}
\usepackage{graphicx,epstopdf}
\usepackage[numbers]{natbib}
\usepackage[makeroom]{cancel}
\usepackage{hyperref}
\usepackage{array}
\usepackage[export]{adjustbox}
\usepackage[normalem]{ulem}
\usepackage{tcolorbox}
\usepackage{romannum}

\title{Spin-induced scalarization of Kerr-Newman black hole  in extended scalar-tensor theory}

\author[a]{Ming Zhang,}
\author[b]{Jie Jiang\footnote{Corresponding author}}

\affiliation[a]{Department of Physics, Jiangxi Normal University,\\ Nanchang 330022, China}
\affiliation[b]{College of Education for the Future, Beijing Normal University, \\ Zhuhai 519087, China}

\emailAdd{mingzhang@jxnu.edu.cn}
\emailAdd{jiejiang@mail.bnu.edu.cn  }

\date{\today}

\abstract{With certain non-minimal coupling between a massless scalar field and the Gauss-Bonnet curvature invariant in the  extended scalar-tensor-Gauss-Bonnet (ESTGB) field theory,  tachyonic instability of Kerr-Newman (KN) black hole is promoted. Critical spin and charge for the onset of the spontaneous scalarization phenomenon for the KN black hole are calculated in the infinitely large coupling limit. Then analytical formula for the coupling-strength-dependent critical existence line is obtained for the black-hole-scalar-field configuration in a finite large coupling regime. Moreover, numerical methods are used to perform threshold curves which are boundaries between bald KN black holes and their hairy counterparts. Intriguingly, we get to know KN black hole can be scalarized in the vanishing spin limit in ESTGB theory.}

\begin{document}

\maketitle

\section{Introduction}
According to the well-known no-hair theorem, an asymptotically flat and stationary black hole spacetime can be characterized at most by three conserved quantities, {\it i.e.,} mass, angular momentum, and electric charge \cite{Chrusciel:2012jk}, which are subject to the Gauss law. What this uniqueness theorem implies is that matter fields propagating in a stationary spacetime will be either absorbed or scattered by the black hole. The theorem, though robust in Einstein's electro-vacuum general relativity (GR), sees some evasion beyond GR or even in GR with typical non-linear matter sources \cite{Herdeiro:2015waa}.  Black holes with additional stable primary or secondary hairs were found, e.g., Refs. \cite{Herdeiro:2014goa,Sotiriou:2015pka,Cardoso:2016ryw,Hod:2016ixl,Wang:2018xhw,Delgado:2020hwr}.

Recently, great progress in investigations of  the destabilization of hair-free  black holes and the occurrence of hairy black holes, dubbed spontaneous scalarization, were made. In fact, earlier in the 1990s, neutron star solution of the scalar-tensor theory was found to be capable of being scalarized induced by scalar field non-minimally coupled with Ricci curvature \cite{Damour:1993hw}.  For the past few years, spontaneous scalarization mechanisms of black holes due to tachyonic instability were disclosed in charged-induced cases \cite{Doneva:2017bvd,Hod:2019pmb,Silva:2017uqg,Antoniou:2017acq,Cunha:2019dwb,Hod:2019vut,Hod:2020jjy}  as well as in curvature-induced cases \cite{Herdeiro:2018wub,Hod:2019ulh,Hod:2020ius,Hod:2020cal}. Furthermore, it was shown that the rotating Kerr black hole can also be scalarized with a specific domain of the spin parameter in the extended scalar-tensor-Gauss-Bonnet (ESTGB) theory, where the electro-vacuum GR is minimally coupled to the Gauss-Bonnet (GB) quadratic curvature invariant through a scalar coupling function $f(\phi)<0$ \cite{Dima:2020yac,Herdeiro:2020wei,Berti:2020kgk}. ESTGB theory is a natural extension of the standard scalar-tensor theory as it is free from ghosts.

Moreover, there are other mechanisms that can result in the spontaneous  scalarization phenomena, such as by magnetic charges \cite{Annulli:2022ivr}, non-linear electromagnetic field \cite{Wang:2020ohb}, and quasi-topological electromagnetism field  \cite{Liu:2019rib,Myung:2020ctt}, in the contexts of different coupling functions \cite{Blazquez-Salcedo:2020nhs}, in  asymptotically (A)dS background \cite{Brihaye:2019gla,Guo:2021zed}, dynamical nonlinear accretion scalarization \cite{Zhang:2021ybj}. One can see the latest review Ref. \cite{Doneva:2022ewd} on spontaneous  scalarization, or even spontaneous vectorization \cite{Minamitsuji:2021rtw,Brihaye:2021qvc} and spontaneous tensorization \cite{Ramazanoglu:2018hwk,Minamitsuji:2020hpl} of both black holes or neutron stars.

Intriguingly, spin-charge induced scalarization of Kerr-Newman (KN) black hole spacetime was studied analytically \cite{Hod:2022txa} and numerically \cite{Lai:2022spn} a few months ago. It was shown that in the Einstein-Maxwell-scalar (EMs) field theory, where the scalar field configuration is non-minimally coupled with the Maxwell electromagnetic invariant of the background spacetime by a negative coupling function, the massless scalar field can be stably supported by the KN black hole. Remarkably, this spontaneous scalarization can only be induced by non-zero spin and charge \cite{Hod:2022txa}.

Natural questions may arise:  can the charged and rotating KN black hole be scalarized by curvature coupling in the ESTGB theory? If the spontaneous scalarization phenomenon occurs, is it induced by only non-zero spin and charge? What is the interplay between the coupling strength, charge, and angular momentum in stimulating tachyonic instability, which is the hallmark of spontaneous scalarization of the KN black hole? In this paper, we will try to clarify these questions and investigate the  tachyonic instability of the Einstein-Maxwell-scalar theory non-minimally coupled with the Gauss-Bonnet (GB) invariant of the background spacetime by a negative coupling function. This exponentially growing instability signifies the onset of the spontaneous scalarization of the bald background spacetime, which we specifically choose the KN one as a prototype. Explicitly, we will investigate the spin-induced scalarization of the KN black hole using both analytic and numerical techniques. To this end, we will first calculate critical angular momentum and charge which mark the boundary between the bald spinning charged KN black hole and its scalarized counterpart in Sec. \ref{sec:onset}. In fact, these  critical values correspond to the case where the coupling strength is infinitely large. We will then derive a compact analytic formula for the existence line in the finitely large coupling regime in Sec. \ref{sec:exi}. In Sec. \ref{sec:naar}, we will numerically study the scalarization of the KN black hole in the ESTGB theory to obtain the whole existence line with arbitrary coupling strength discerning bald and scalarized KN black holes and confirm our  analytical results in former sections. The last section will be devoted to our concluding remarks.

\section{Onset of spontaneous scalarization of Kerr-Newman black hole}\label{sec:onset}
In this section, we will investigate the onset of spontaneous scalarization of the KN black hole in the  ESTGB field theory, which can be described by the action
\begin{equation}\label{act}
S=\int d^4 x \sqrt{-g}\left[R-F_{\mu \nu} F^{\mu \nu}-2 \nabla_\mu \phi \nabla^\mu \phi+f(\phi) R_{\mathrm{GB}}^2\right],
\end{equation}
where $R$ is the Ricci scalar, $F_{\mu \nu}$ is the electromagnetic strength tensor, $\phi$ is a real scalar field, $f(\phi)$ is some coupling function controlling the  non-minimal coupling between the scalar field and the GB invariant which explicitly reads
\begin{equation}
R_{\mathrm{GB}}^2=R_{\mu \nu \rho \sigma} R^{\mu \nu \rho \sigma}-4 R_{\mu \nu} R^{\mu \nu}+R^2.
\end{equation}
We shall show that the non-trivial coupling between the scalar function $f(\phi)$ and the GB curvature term yields spontaneous scalarization of the KN black hole, which can be viewed as a tachyonic instability at a linear level. To this end, we choose the coupling function to be a universal quadratic form
\begin{equation}
f(\phi)=-\frac{1}{8}\lambda \phi^2,
\end{equation}
where the coupling strength parameter $\lambda$ is dimensionful. It guarantees that the bald KN black hole is an electrovacuum solution of the ESTGB theory in the limit $\phi\to 0$. The coupling parameter can be either positive or negative. To be specific, we will focus on the case $\lambda>0$ in what follows. In the terminology of Ref. \cite{Herdeiro:2021vjo}, this case corresponds to the $\mathrm{GB}^{-}$ scalarization. We will show that the KN black hole can support linearized spatially regular scalar configurations in case of  a negative coupling between the GB invariant of the background spacetime and the scalar field.

The action (\ref{act}) admits a bald KN spacetime solution 
\begin{equation}
\begin{aligned}
d s^2=&-\frac{\Delta}{\rho^2}\left(d t-a \sin ^2 \theta d \varphi\right)^2+\frac{\rho^2}{\Delta} d r^2+\rho^2 d \theta^2\\&+\frac{\sin ^2 \theta}{\rho^2}\left[a d t-\left(r^2+a^2\right) d \varphi\right]^2
\end{aligned}
\end{equation}
 in the vanishing scalar field limit, where $\Delta = r^2-2 M r+a^2+Q^2$, $\rho^2 = r^2+a^2 \cos ^2 \theta$. $M,\,a\,,Q$ are respectively the mass, spin, and electric charge of the black hole. When $a=0$, we obtain the Reissner-Nordstr\"{o}m (RN) black hole. The event horizon locates at $r_{\pm}=M \pm\left(M^2-a^2-Q^2\right)^{1 / 2}$. The equation of motion for the scalar field derived from the action (\ref{act})  is given by the Klein-Gordon (KG) equation
\begin{equation}\label{kgeq}
\left(\nabla^\mu \nabla_\mu-\mu_{\mathrm{eff}}^2\right) \phi=0,
\end{equation}
where the effective mass is explicitly given by
\begin{equation}
\mu_{\mathrm{eff}}^2(r, \theta, M, Q, a)=\frac{\lambda}{4} \cdot R_{\mathrm{GB}}^2(r, \theta, M, Q, a),
\end{equation}
with
\begin{equation}
\begin{aligned}
\frac{r^8}{Q^4}R_{\mathrm{GB}}^2=&\frac{8}{\bar{y}^2 (1+\bar{x})^6}\left[-6 \bar{x}^6+5 \bar{y}^2-12 \bar{y}+6\right.\\&\left.+5 \bar{x}^4 \left(\bar{y}^2-12 \bar{y}+18\right)\right.\\&\left.-2 \bar{x}^2 \left(19 \bar{y}^2-60 \bar{y}+45\right)\right],
\end{aligned}
\end{equation}
where we have defined two dimensionless quantities
\begin{equation}\label{xbyb}
\bar{x} = \frac{a \cos \theta}{r},\quad \bar{y}=\frac{Q^2}{Mr}.
\end{equation}
Not loss of generality, we can choose $a\geq 0$. It is evident that
\begin{equation}
0\leq\bar{x}\le \frac{a^2}{r^2}\le\frac{a^2}{r_+^2}\le\frac{a^2}{M^2}\le 1,
\end{equation}
\begin{equation}\label{yra}
0\le \bar{y}\le \frac{Q^2}{M^2}\le 1.
\end{equation}

To expand the  KG equation (\ref{kgeq}), we can define a  tortoise coordinate $x$ by
\begin{equation}\label{tort}
\frac{d x}{d r}=\frac{r^2+a^2}{\Delta},
\end{equation}
and introduce an alternative azimuthal coordinate $\bar{\varphi}$ via 
\begin{equation}
d \bar{\varphi}=d \varphi+\frac{a}{\Delta} d r.
\end{equation}
As a result, the problem of the Boyer-Lindquist coordinate $\varphi$ being ill-defined on the horizon is resolved.
Then we separate  the azimuthal dependence of the scalar field $\phi$ by
\begin{equation}
\phi=\frac{1}{r} \Psi(t, r, \theta) e^{i m \bar{\varphi}},
\end{equation}
where $m$ is an azimuthal harmonic index for perturbation modes of the scalar field.
This decomposition leads to a (2+1)-dimensional equation
\begin{equation}\label{tpoeq}
\begin{aligned}
&\left\{-\Sigma^2 \partial_{t t}+\left(r^2+a^2\right)^2 \partial_{x x}+2 i a m \left( Q^2-2 M r\right) \partial_t\right. \\
&\left.\quad+\left[2 i a m\left(r^2+a^2\right)-2 a^2 \Delta / r\right] \partial_x-\hat{V}\right\} \Psi=0,
\end{aligned}
\end{equation}
where
\begin{equation}
\Sigma^2=\left[\left(r^2+a^2\right)^2-a^2 \Delta\right]+a^2 \Delta \cos ^2 \theta,
\end{equation}
\begin{equation}
\begin{aligned}
\hat{V}=& \Delta\left[-\left\{\partial_{\theta \theta}+\cot \theta \partial_\theta-\frac{m^2}{\sin ^2 \theta}\right\}+\frac{2 M}{r}\left(1-\frac{a^2+Q^2}{M r}\right)\right.\\
&\left.+\frac{2 i a m}{r}+\left(r^2+a^2 \cos ^2 \theta\right) \mu_\mathrm{eff}^2\right] .
\end{aligned}
\end{equation}
We will use this equation to numerically solve the scalar field in Sec. \ref{sec:naar}. 
On the other hand, as suggested in Refs. \cite{Dima:2020yac,Dolan:2012yt}, we can project the Eq. (\ref{kgeq}) onto a basis of spherical harmonic functions $Y_{j m}(\theta)$, yielding  evolution equations in coordinates $t$ and $r$ for the components of the scalar field $\psi_{l m}(t, r)$,
\begin{equation}\label{wyjm}
\begin{aligned}
&{\left[\left(r^2+a^2\right)^2-a^2 \Delta\left(1-c_{l l}^m\right)\right] \ddot{\psi}_{l m}} \\
&\quad+a^2 \Delta\left(c_{l, l+2}^m \ddot{\psi}_{l+2, m}+c_{l, l-2}^m \ddot{\psi}_{l-2, m}\right) \\
&\quad+2 i a m (2M r-Q^2) \dot{\psi}_{l m}-\left(r^2+a^2\right)^2 \psi_{l m}^{\prime \prime} \\
&\quad-\left[2 i a m\left(r^2+a^2\right)-2 a^2 \Delta / r\right] \psi_{l m}^{\prime} \\
&\quad+\Delta\left[l(l+1)+2 M / r-2(a^2+Q^2) / r^2+2 i a m / r\right] \psi_{l m} \\
&\quad+\Delta \sum_j\left\langle l m\left|\mu_{\mathrm{eff}}^2\left(r^2+a^2 \cos ^2 \theta\right)\right| j m\right\rangle \psi_{j m}=0,
\end{aligned}
\end{equation}
where we have denoted
\begin{align}
c_{j l}^m &=\left\langle l m\left|\cos ^2 \theta\right| j m\right\rangle,\\
\psi_{l m}(t, r) &= \int r \phi Y_{l m} Y_{l m}^* d \Omega \equiv\langle lm|r \phi| lm\rangle.
\end{align}
${}^\prime$ and $\dot{}$ denote derivative with respect to $r$ and $t$, respectively. $l, j$ are spheroidal harmonic indexes for perturbation modes of the scalar field.

According to the  illustration in \cite{Dima:2020yac}, the main tachyonic instability of the black hole-scalar field system comes from the $m=0$ mode with an infinitely long instability timescale, so at the late time we have
\begin{align}
 &\quad\sum_j\left\langle l m\left|\mu_{\mathrm{eff}}^2\left(r^2+a^2 \cos ^2 \theta\right)\right| j m\right\rangle \psi_{j m}\\&=
\left\langle l_1 0\left|\mu_{\mathrm{eff}}^2\left(r^2+a^2 \cos ^2 \theta\right)\right| l_2 0\right\rangle \psi_{l_2 0}.
\end{align}

\begin{figure}[htpb!]
\begin{center}
\includegraphics[width=3.3in,angle=0]{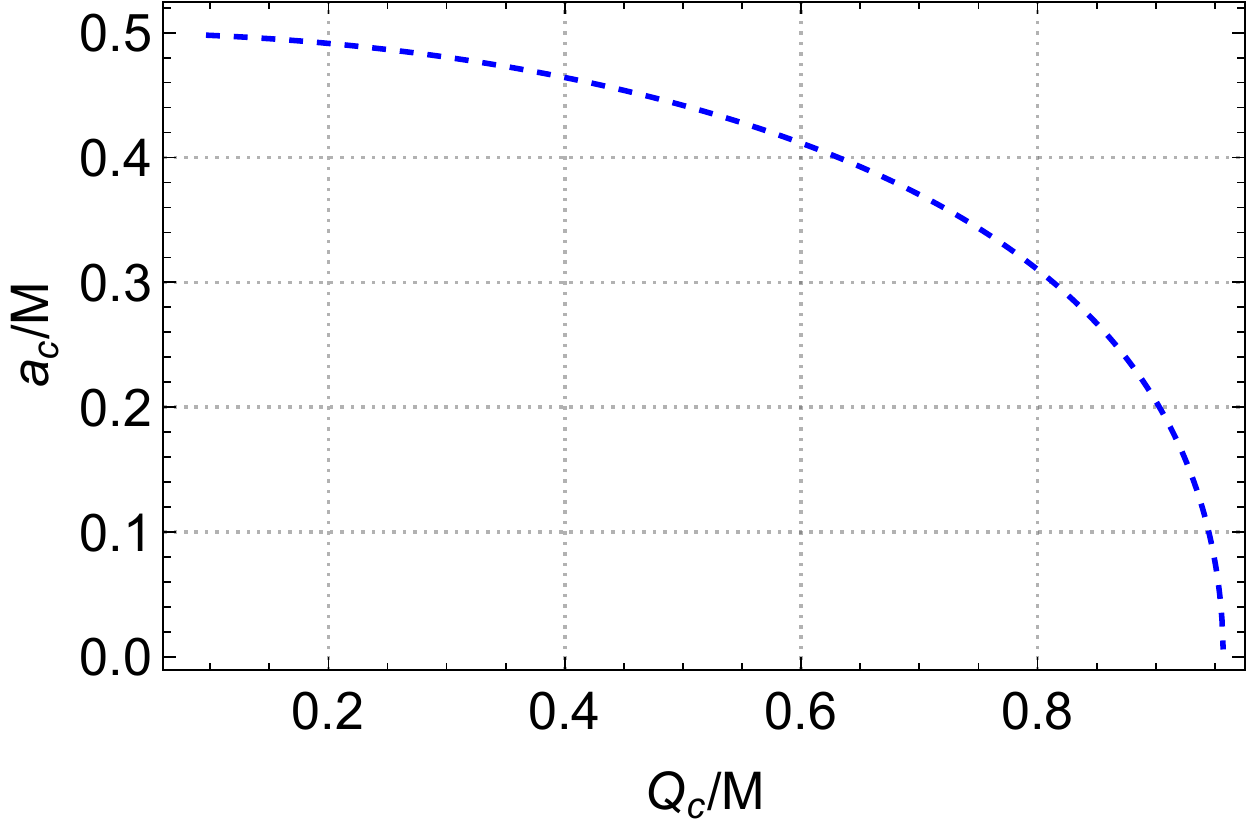}
\end{center}
\vspace{-5mm}
 \caption {Variation of the critical electric charge $Q_c$ with respect to the critical angular momentum $a_c$ which marks the onset of the scalarization of the KN black hole in the unit of black hole mass $M$.}\label{aqm}
\end{figure}

Moreover, the spontaneous scalarization of the black hole-scalar field system is closely related to a negative effective squared mass term (or in other words, a negative binding potential well) nearby the black hole's event horizon, with a relation $r_{\text {out }} \geq r_{\text {in }}=r_{+}$, where $r_{\mathrm{in}},  r_{\text {out }}$ are two turning points of the effective potential. Specifically, the boundary between the bald KN black hole and  scalarized black hole corresponds to a degenerate effective potential well characterized by two turning points merging at the event horizon of the black hole. Similar to the  Kerr case in Ref. \cite{Dima:2020yac} (as will be proved numerically in Sec. \ref{sec:naar}), the boundary is characterized by a coupling parameter of $\lambda\to\infty$, which results in that all terms except the last one in Eq. (\ref{wyjm}) can be neglected. That is, for some critical angular momentum $a_c$ and electric charge $Q_c$, we have
\begin{equation}\label{inout}
\left\langle l_1 0\left|\mu_{\mathrm{eff}}^2\left(r^2+a^2 \cos ^2 \theta\right)\right| l_2 0\right\rangle_{r=r_{\text {in }}=r_{\text {out }}=r_{+}\left(a_c, Q_c\right)}=0,
\end{equation}
which is equivalent to
\begin{equation}
\begin{aligned}
\int_0^\pi & \mu_{\mathrm{eff}}^2 \left(r^2+a^2 \cos ^2 \theta\right)  Y_{l_1 m^*=0}(\cos \theta) \\& \cdot Y_{l_2 m^*=0}(\cos \theta) \sin \theta d \theta=0.
\end{aligned}
\end{equation}
As $Y_{l 0}(\cos \theta)$ becomes gradually close to a delta function peaking at $\theta=0, \pi$ in the regime $l_1=l_2 \rightarrow \infty$, we  obtain
\begin{equation}\label{bblef}
\begin{aligned}
6 \bar{x}^6-5 \bar{x}^4 \left(\bar{y}^2-12 \bar{y}+18\right)&+2 \bar{x}^2 \left(19 \bar{y}^2-60 \bar{y}+45\right)\\&-5 \bar{y}^2+12 \bar{y}-6=0.
\end{aligned}
\end{equation}
Taking the restriction (\ref{yra}) into account, we have
\begin{equation}
\begin{aligned}
\bar{y}=&\frac{1}{5 \bar{x}^4-38 \bar{x}^2+5}\left(30 \bar{x}^4-60 \bar{x}^2+6\right.\\&\left. -\sqrt{6} \left(\bar{x}^2+1\right) \sqrt{5 \bar{x}^6+27 \bar{x}^4-9 \bar{x}^2+1}\right).
\end{aligned}
\end{equation}
So for any given critical value of $\bar{x}_c\in [0,1]$, we can obtain the other critical of  $\bar{y}_c\in [0,1]$ by using the above expression. Then according to Eq. (\ref{xbyb}), we have 
\begin{align}\label{xbard}
\bar{x}_c=\frac{a_c}{r_+},\quad \bar{y}_c=\frac{Q^2_c}{Mr_+}.
\end{align}
Consequently, the critical angular momentum $a_c$ and electric charge $Q_c$ that signify the beginning of the scalarization for the KN black hole can be calculated. For instance: for $\bar{y}=1/5$, we get $a_c=0.41748 M,\,Q_c=0.58263 M$; for $\bar{y}=7/10$, we have $a_c=0.05859 M,\,Q_c=0.95297 M$. All values of $Q_c$ and $a_c$ form an existence line for the infinitely large coupling case $\lambda\to \infty$. To facilitate comparing between analytical and numerical results, we plot the relation between $a_c$ and $Q_c$ in Fig. \ref{aqm}. Evidently, we have $a_c= 0.5 M$ for $Q_c= 0$, which is in accordance with the result obtained numerically in Ref. \cite{Dima:2020yac} and analytically in Ref. \cite{Hod:2020jjy}. At the same time, we have  $Q_c\to 0.957 M$ in the limit $a_c\to 0$. This is intriguing as it means that the charged  RN black hole in the ESTGB theory can be scalarized in the highly charged regime.   Similar results are found in Refs. \cite{Herdeiro:2018wub,Fernandes:2019rez,Hod:2019ulh}, where the onset of  spontaneous scalarization of the RN black hole was found in the EMs theory.

\section{Existence line in the large coupling regime}\label{sec:exi}
In the last section, we obtained the critical values of angular momentum $a_c$ and electric charge $Q_c$ characterizing the  onset of the scalarization of the KN black hole in the infinitely large coupling regime $\lambda\to\infty$ for the ESTGB theory.  In what follows, we will calculate the existence line $\bar{\lambda}=\bar{\lambda}(\bar{a}, \bar{Q})$ for the  black hole-scalar field system in the large  coupling regime $\bar{\lambda}\gg 1$, where we have defined
\begin{align}\label{redu}
\bar{\lambda}=\frac{\lambda}{M^2},\,\bar{a}=\frac{a}{M},\,\bar{Q}=\frac{Q}{M}.
\end{align}
The line gives the boundary between the bald KN black hole and the scalarized KN black hole supporting spatially regular hairy scalar configurations in the ESTGB theory and it exhibits constraint condition between the reduced angular momentum $\bar{a}$, reduced electric charge $\bar{Q}$, and the reduced coupling parameter $\bar{\lambda}$ of  the black hole-scalar field system in the large (but finite) coupling regime.
To this end, we  substitute the KN metric into the KG equation (\ref{kgeq}) and obtain 
\begin{equation}\label{kgeq2}
\begin{aligned}
&\left\{\frac{1}{\Delta}\left[\left(r^2+a^2\right)^2-\Delta a^2 \sin ^2 \theta\right] \frac{\partial^2}{\partial t^2}\right.\\&\left.-\frac{\partial}{\partial r}\left(\Delta \frac{\partial}{\partial r}\right)-\frac{1}{\sin \theta} \frac{\partial}{\partial \theta}\left(\sin \theta \frac{\partial}{\partial \theta}\right)\right. \\
&\left.-\frac{1}{\Delta \sin ^2 \theta}\left(\Delta-a^2 \sin ^2 \theta\right) \frac{\partial^2}{\partial \varphi^2}-\frac{2a\left( 2Mr-Q^2 \right)}{\Delta} \frac{\partial^2}{\partial t \partial \varphi}\right.\\&\left.+\left(r^2+a^2 \cos ^2 \theta\right) \mu_{\mathrm{eff}}^2(r, \theta)\right\} \phi=0.
\end{aligned}
\end{equation}
Then we substitute a decomposition  \cite{Hod:2022hfm}
\begin{equation}
\phi(r, \theta, \varphi)=\sum_j R_j(r) Y_{j 0}(\theta)
\end{equation}
 for the static massive scalar field with zero-frequency into Eq. (\ref{kgeq2}) and multiply it with $Y_{l 0}^*$ before integrating $\theta$ and $\phi$, resulting in a simpler form
\begin{equation}\label{eomphi}
\begin{aligned}
&\frac{d}{d r}\left(\Delta \frac{d R_l}{d r}\right)-l(l+1) R_l\\&-\sum_j R_j(r) \int_0^{2 \pi} \int_0^\pi\left(r^2+a^2 \cos ^2 \theta\right)\\&\quad\quad\quad\quad\quad\quad \cdot \mu_{\mathrm{eff}}^2(r, \theta) Y_{l 0}^* Y_{j 0} \sin \theta d \theta d \varphi=0.
\end{aligned}
\end{equation}

In the near-critical regime $\bar{a}\sim \bar{a}_c\,,\bar{Q}\sim \bar{Q}_c$, with $ \bar{a}_c\equiv a_c/M,\,\bar{Q}_c\equiv Q_c/M$, the hairy KN black hole is characterized by the large coupling $\bar{\lambda}\gg 1$. We shall consider a marginally stable scalar mode with an infinitely long instability timescale. In this regime, other modes decay in time, leaving Eq. (\ref{eomphi}) to be
\begin{equation}
\begin{gathered}
\frac{d}{d r}\left(\Delta \frac{d R_l}{d r}\right)-l(l+1) R_l-R_l \int_0^{2 \pi} \int_0^\pi\left(r^2+a^2 \cos ^2 \theta\right) \\
\cdot \mu_{\mathrm{eff}}^2(r, \theta) Y_{l 0}^* Y_{l 0} \sin \theta d \theta d \varphi=0,
\end{gathered}
\end{equation}
which can be further simplified as
\begin{equation}
\begin{aligned}
\frac{d}{d r} &\left(\Delta \frac{d R_l}{d r}\right)+\left[\frac{\left(r^2+a^2 \right)\mu_{\mathrm{eff}}^2}{2\pi}-l(l+1)\right] R_l=0
\end{aligned}
\end{equation}
after considering asymptotic eikonal limit $l\gg 1$.

We can redefine the scalar field by
\begin{equation}
\Upsilon \equiv r R
\end{equation}
and use a new coordinate
\begin{equation}\label{tcx}
d r_*=\frac{r^2}{\Delta} \cdot d r.
\end{equation}
Then we obtain a Schrödinger-like differential equation
\begin{equation}\label{slde}
\frac{d^2 \Upsilon}{d r_*^2}-V\Upsilon=0,
\end{equation}
where the effective potential can be split into two parts,
\begin{equation}
V(r)=V_l (r)+V_{\mathrm{GB}}(r)
\end{equation}
which individually reads
\begin{equation}\label{eq:m2vx}
\begin{aligned}
M^2 V_l(r)&=\frac{\Delta \left(\Delta '+l (l+1) r\right)}{r^5}-\frac{2 \Delta ^2}{r^6}\\
&=\frac{M^2}{r^2}\left(-\frac{2 a^2}{r^2}+l^2+l+\frac{2 M}{r}-\frac{2 Q^2}{r^2}\right)h(r),
\end{aligned}
\end{equation}
\begin{equation}
\begin{aligned}
&M^2 V_{\mathrm{GB}}(r)=\frac{\Delta  \mu_{\mathrm{eff}}^2  \left(a^2+r^2\right)}{2 \pi  r^4}\\
=&-\frac{ \bar{\lambda}  M^6}{\pi  r^6}\times \frac{1}{\left(\frac{a^2}{r^2}+1\right)^5}\times\left[\frac{6 a^6}{r^6}-6-\frac{5 a^4 Q^4}{M^2 r^6}+\frac{60 a^4 Q^2}{M r^5}\right.\\&\left.    -\frac{90 a^4}{r^4}+\frac{38 a^2 Q^4}{M^2 r^4}-\frac{120 a^2 Q^2}{M r^3}+\frac{90 a^2}{r^2}\frac{12 Q^2}{M r}-\frac{5 Q^4}{M^2 r^2}\right.\\&\left.+\frac{12 Q^2}{M r}\right].
\end{aligned}
\end{equation}
Note that we have defined 
\begin{equation}
h(r) \equiv \frac{\Delta}{r^2}=1-\frac{2 M}{r}+\frac{a^2}{r^2}+\frac{Q^2}{r^2}
\end{equation}
in Eq. (\ref{eq:m2vx}).

Using the WKB method \cite{bender1999advanced}, we have the discrete quantization condition for the Schrödinger-like differential equation (\ref{slde}),
\begin{equation}\label{wkb1}
\begin{aligned}
\int_{r_{*-}}^{r_{*+}} d r_* \sqrt{-V(r_*, M, a, Q, \bar{\lambda})} =\left(n+\frac{1}{2}\right) \pi,
\end{aligned}
\end{equation}
where $n=0, 1, 2, \cdots$ is the overtone number and $\left\{r_{*-}, r_{*+}\right\}$  are the classical turning points of the effective binding potential function $V$ which satisfies $V\left(r_{*-}\right)=V\left(r_{*+}\right)=0$. The existence line is determined by the fundamental mode $n=0$. According to Eq. (\ref{tcx}), the WKB resonance equation (\ref{wkb1}) can be rewritten as
\begin{equation}\label{wkb2}
\begin{aligned}
\int_{r_{t-}}^{r_{t+}} dr \frac{\sqrt{-V(r, M, a, \bar{\lambda})}}{h(r)} &=\left(n+\frac{1}{2}\right)  \pi.
\end{aligned}
\end{equation}

As we aim to attain the existence line between the bald and scalarized KN black hole in the large coupling regime $\bar{\lambda}\gg 1$, which corresponds to the near-critical regime $\bar{a}\sim \bar{a}_c\,,\bar{Q}\sim \bar{Q}_c$, it turns out to be convenient to introduce some dimensionless parameters $\epsilon,\,\chi\,,$ and $\sigma$, which are defined by
\begin{align}
a &\equiv a_{c} \cdot(1+\epsilon)\label{amx1},\\
Q &\equiv Q_{c} \cdot(1+\chi),\\
r &\equiv r_{+} \cdot(1+\sigma)\label{amx3}.
\end{align}
As a result, we can respectively express $r/M,\,a/r\,,Q/r$  as
\begin{equation}\label{rm1}
\begin{aligned}
\frac{r}{M}&=\frac{-a_c^2 (\sigma+\epsilon +1)-Q_c^2 (\chi +\sigma+1)+(\tau +1) (\sigma+1)}{\tau }\\&\quad+\mathcal{O}(\sigma^2, \epsilon^2, \chi^2, \sigma\epsilon, \sigma\chi, \epsilon\chi),
\end{aligned}
\end{equation}
\begin{equation}
\begin{aligned}
\frac{a}{r}&=\frac{2 a_c (-\sigma +\epsilon +1)}{\tau  (\tau +1)^2}-\frac{a_c^3 (\epsilon -(\sigma -1) (\tau +2))}{\tau  (\tau +1)^3}\\&\quad+\frac{a_c Q_c^2 (\sigma  (\tau +2)+\tau  \chi -\tau +\chi -(\tau +2) \epsilon -2)}{\tau  (\tau +1)^3}\\&\quad+\mathcal{O}(\sigma^2, \epsilon^2, \chi^2, \sigma\epsilon, \sigma\chi, \epsilon\chi),
\end{aligned}
\end{equation}
\begin{equation}
\begin{aligned}
\frac{Q}{r}&=\frac{a_c^2 Q_c ((\tau +2) (\sigma -\chi -1)+(\tau +1) \epsilon )}{\tau  (\tau +1)^3}\\
&\quad-\frac{2 Q_c (\sigma -\chi -1)}{\tau  (\tau +1)^2}+\frac{Q_c^3 (\sigma  (\tau +2)-\tau -\chi -2)}{\tau  (\tau +1)^3}\\&\quad+\mathcal{O}(\sigma^2, \epsilon^2, \chi^2, \sigma\epsilon, \sigma\chi, \epsilon\chi).
\end{aligned}
\end{equation}
where we have denoted 
\begin{equation}\label{taudef}
\tau=\sqrt{1-Q_c^2-a_c^2}.
\end{equation}
Besides, near the horizon, we have
\begin{equation}\label{hor}
h(r)=\varpi_1 \cdot \sigma+\mathcal{O}\left(x^2, \epsilon^2, \chi^2, \sigma \epsilon, \sigma \chi, \epsilon\chi\right)
\end{equation}
with $\varpi_1=\varpi_1 (a_c, Q_c)>0$ a dimensionless number though its explicit form is lengthy. Intriguingly, it is independent of the angular momentum and electric charge of the black hole.

Using Eqs. (\ref{rm1})-(\ref{hor}), we obtain
\begin{equation}
M^2 \frac{V(r)}{\left[h(r)\right]^2}=\frac{\bar{\lambda}}{4}\left(\frac{\varpi_2}{\varpi_1} -\frac{\varpi_3 \epsilon+\varpi_4 \chi}{\varpi_1\sigma}\right),
\end{equation}
where $\varpi_2=\varpi_2 (a_c, Q_c)>0, \varpi_3=\varpi_3 (a_c, Q_c)>0, \varpi_4=\varpi_4 (a_c, Q_c)>0$ are dimensionless numbers which are quite cumbersome. Then the WKB resonance equation (\ref{wkb2}) becomes
\begin{equation}\label{rmf}
\begin{aligned}
&\frac{r_{+}}{M} \cdot \int_0^{\frac{\varpi_1}{\varpi_2}\cdot\frac{\varpi_3\epsilon+\varpi_4\chi}{\varpi_1}} d\sigma \sqrt{\frac{\bar{\lambda}}{4} \cdot \frac{\varpi_2}{\varpi_1} \cdot\left(\frac{\varpi_1}{\varpi_2}\cdot\frac{\varpi_3\epsilon+\varpi_4\chi}{\varpi_1 \sigma}-1\right)} \\
&=\left(n+\frac{1}{2}\right) \pi.
\end{aligned}
\end{equation}
To work out this integral, we can introduce a dimensionless quantity $z$ by
\begin{equation}
z=\frac{\varpi_1}{\varpi_3\epsilon+\varpi_4\chi}.
\end{equation}
Then Eq. (\ref{rmf}) becomes
\begin{equation}
\begin{aligned}
&\frac{r_{+}}{M} \cdot \sqrt{\frac{\bar{\lambda}}{4} \cdot \frac{\varpi_2}{\varpi_1}}\cdot \frac{\varpi_3\epsilon+\varpi_4\chi}{\varpi_1}\cdot\frac{\varpi_1}{\varpi_2}\int_0^{1} dz \sqrt{\frac{1}{z}-1} \\
&=\left(n+\frac{1}{2}\right)  \pi.
\end{aligned}
\end{equation}
Subsequently, we get an explicit form of the discrete resonance spectrum equation of the KN black hole-scalar field system in the large coupling regime, which reads
\begin{equation}\label{reno}
\bar{\lambda}=\frac{16 \varpi_1 \varpi_2}{\left(1+\tau\right)^2\left(\varpi_3 \epsilon+\varpi_4 \chi\right)^2}\left(n+\frac{1}{2}\right)^2
\end{equation}
Using Eqs. (\ref{amx1})-(\ref{amx3}), the above relation can be  equivalently expressed as
\begin{equation}\label{exlal}
\varpi_3 \frac{\bar{a}}{a_c}+\varpi_4 \frac{\bar{Q}}{Q_c}=\varpi_3+\varpi_4+\frac{4\left(n+\frac{1}{2}\right)}{1+\tau} \sqrt{\frac{\varpi_1 \varpi_2}{\bar{\lambda}}}.
\end{equation}
The critical existence line can be extracted from the above resonance formula (\ref{reno}) by considering the fundamental mode $n=0$, 
\begin{equation}
\sqrt{\bar{\lambda}}=\frac{2\sqrt{\varpi_1 \varpi_2}}{\varpi_3 \left(\frac{\bar{a}}{a_c}-1\right)+\varpi_4 \left(\frac{\bar{Q}}{Q_c}-1\right)}\cdot\frac{1}{1+\tau},
\end{equation}
where $\tau$ has been defined in Eq. (\ref{taudef}). This compact characterized line formula marks the boundary between the hairy charged spinning black hole and the bald KN black hole in the large coupling regime $\bar{\lambda}\gg 1$. Evidently, it means that  the  value of $a$ (or $Q$) on the existence line decreases with increasing $\lambda$ for fixed $Q$ (or $a$). The logic here is that, for instance, with a given electric charge $Q$, which can be viewed as $Q_c$, we can first calculate the critical value of the spin $a_c$ in the infinitely large coupling limit. Then a perturbation can be conducted based on those values and an existence line can be obtained. values of spin and charge on that line denote critical values of the onset of the spontaneous scalarization in the finitely large coupling regime. In the vanishing electric charge limit, the result reduces to the one in Ref. \cite{Hod:2022hfm}.

\section{Numerical approach and result}\label{sec:naar}
We are now to numerically solve the (2+1)-dimensional partial differential equation (\ref{tpoeq}) with some properly chosen boundary and initial conditions. The approach makes us directly evolve the scalar wave function  in time so that we can determine the threshold for the KN black hole's destabilization. The angular boundary condition is imposed to ensure smoothness of the solution at the rotation axis ( $\theta=0,\,\pi$), {\it i.e.}, the scalar field must be vanishing at the rotation axis for $m>0$, and its $\theta$-derivative at the rotation axis has to be zero for $m=0$. At the radial direction, the ingoing (outgoing) boundary condition at the horizon (infinity) should be imposed. However, the computational domain is always finite and its edge can reach neither the horizon nor the infinity, which will lead to undesired reflections from the boundaries. To overcome this problem, we push the outer boundary sufficiently far away from the location where the signal is extracted. Under this setup, the radial boundary conditions will not affect our numerical result. 

We choose Gaussian initial conditions
\begin{equation}\label{initialcd}
\Psi(t=0, x, \theta) \sim Y_{l m}(\theta) \exp \left[\frac{-\left(x-x_c\right)^2}{2 \sigma^2}\right]
\end{equation}
centered at $x_c$ for each mode, with $Y_{l m}(\theta)$  a $\theta$-dependent spherical harmonics and $\sigma$ the width of the Gaussian bell.

It is worthy noting that the coefficients of the field equation (\ref{tpoeq}) become singular at the rotation axis. To eliminate the singularity of the equation, we introduce
\begin{equation}
\Psi(t, x, \theta)= \sin^m \theta \bar{\psi}(t, x, \theta).
\end{equation}
After substituting the above equation into Eq. (\ref{tpoeq}) and multiplying both sides by $\sin^{-m} \theta$, we get a regular equation about $\bar{\psi} (t, x, \theta)$ at the rotation axis. From the field equation Eq.  (\ref{wyjm}), we can see that the even-$l$ and odd-$l$ modes decouple, {\it i.e.}, the time-evolution of a mode $(l, m)$ is only a combination of all the modes $(l+2k, m)$ with $k$ some integer. Choosing the initial condition Eq. (\ref{initialcd}), the property means that $\bar{\psi}(t, x, \theta)=\tilde{\psi}(t, x, y)$ for even $(l-m)$ and  $\bar{\psi}(t, x, \theta)=\cos\theta \tilde{\psi}(t, x, y)$ for odd $(l-m)$, in which we define $y=\cos^2\theta$. Using the new variable $y$, Eq. (\ref{tpoeq}) can be written as an equation about $\tilde{\psi}(t, x, y)$. Then it is not hard to verify coefficients of the field equation about $\tilde{\psi}(t, x, y)$ is analytic with respect to $y$ at the domain $[0, 1]$ and thus $\tilde{\psi}(t, x, y)$ is also analytic. Then we can use the pseudospectral method to solve the $y$-differential equation, i.e., we can do an expansion
\begin{equation}
\tilde{\psi}(t, x, y)=\sum_{i=0}^{N}\tilde{\psi}_i(t, x)C_i(y)\,,
\end{equation}
in which
\begin{equation}\begin{aligned}
&C_i(y)=\frac{2}{Np_i}\sum^N_{m=0}\frac{1}{p_m}T_m(1-2y_i) T_m(1-2y)\,,\\
&p_0=p_N=2\,,\quad p_j=1\,,\quad\quad (j=1,2\dots,N-1)
\end{aligned}\end{equation}
with $y_i=1-\cos(i\pi/N)$ is an $N$th-order cardinal function formed by the Chebyshev polynomial $T_n(x)$ and it satisfies $C_i (y_j)=\delta_{ij}$. Then, it is straightforward to show
\begin{equation}\begin{aligned}
\tilde{\psi}_i(t, x)&=\tilde{\psi}(t, x, y_i)\,,\\
\partial_y \tilde{\psi}(t, x, y_i)&=\sum^N_{i=0}\sum^N_{j=0} C'_i(y_j)\tilde{\psi}_i(t, x) C_j(y)\,,\\
\partial_y^2 \tilde{\psi}(t, x, y_i)&=\sum^N_{i=0}\sum^N_{j=0} C''_i(y_j)\tilde{\psi}_i(t, x) C_j(y)\,.
\end{aligned}\end{equation}
With these in mind, we can expand the field equation by a cardinal function and it becomes $(N+1)$ equations about $\tilde{\psi}_i(t, x)$ with $i=0, \cdots, N$. To evolve the equation (\ref{tpoeq}), we also apply the finite difference method to calculate the $x$-direction, and the time integration is conducted by a fourth-order Runge-Kutta method, in which we introduce an auxiliary field $\Pi(t, x, y)=\partial_t \tilde{\psi}(t, x, y)$ to decompose the field equation into two coupled equations with only first-order time derivative. 

To ensure that our result is reliable, we also use the numerical method to solve the field equation (\ref{tpoeq}) about $\Psi(t, x, \theta)$ by applying the finite difference method at $x$- and $\theta$-directions and Runge-Kutta method at time-direction. The corresponding boundary conditions at the rotation axial are discussed in the first paragraph of this section. We will denote this method and the pseudospectral method at $y$-direction as ``FFR'' and ``FPR'' methods individually.

\begin{figure*}[htpb!]
\begin{center}
\includegraphics[width=3.5in,angle=0]{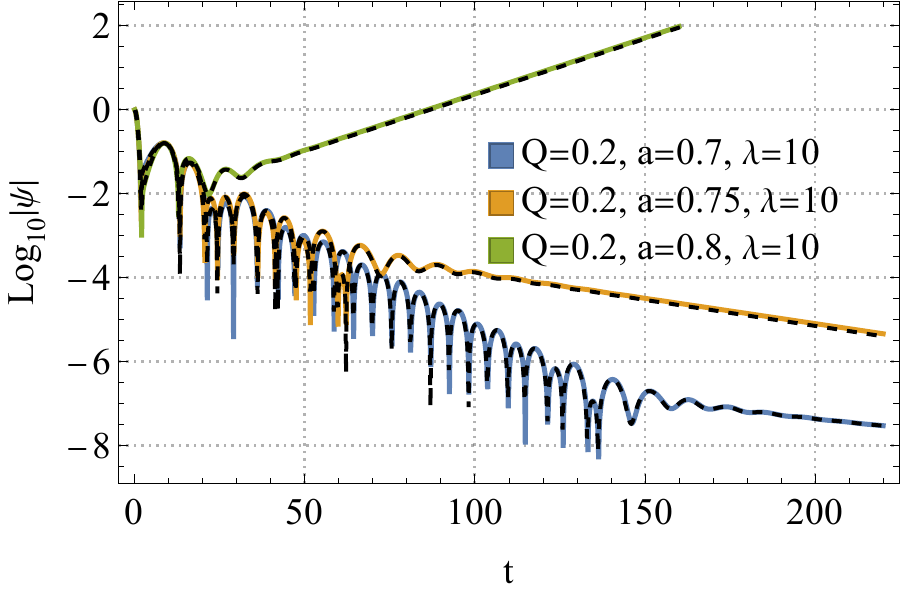}
\includegraphics[width=3.5in,angle=0]{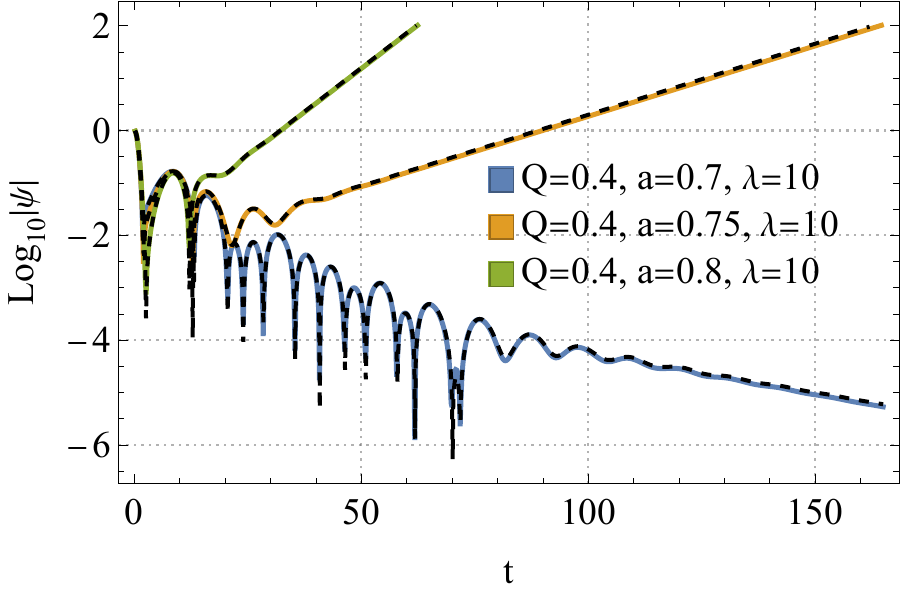}
\end{center}
\vspace{-5mm}
 \caption {The time evolution of perturbation of the scalar field for $M=1$ obtained via FFR method (dashed lines) and FPR method (solid lines) with different coupling constant $\lambda$. The initial mode is chosen to be the one with $l=m=0$. One can see that the same behaviors of the massless scalar field's late-time tails are shared between results via both ways.}\label{eld}
\end{figure*}

\begin{figure*}[htpb!]
\begin{center}
\includegraphics[width=3in,angle=0]{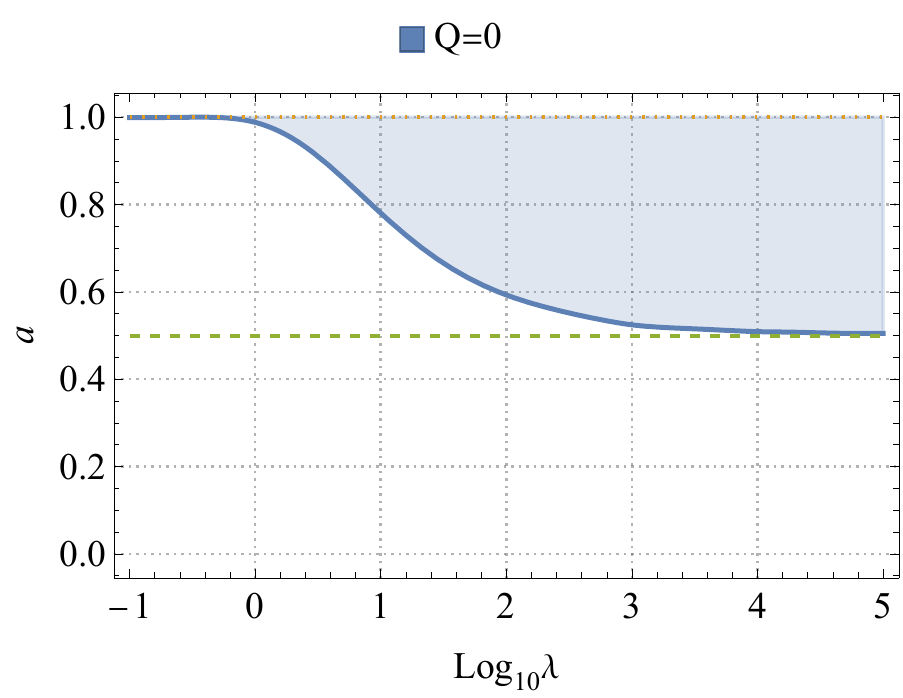}
\includegraphics[width=3in,angle=0]{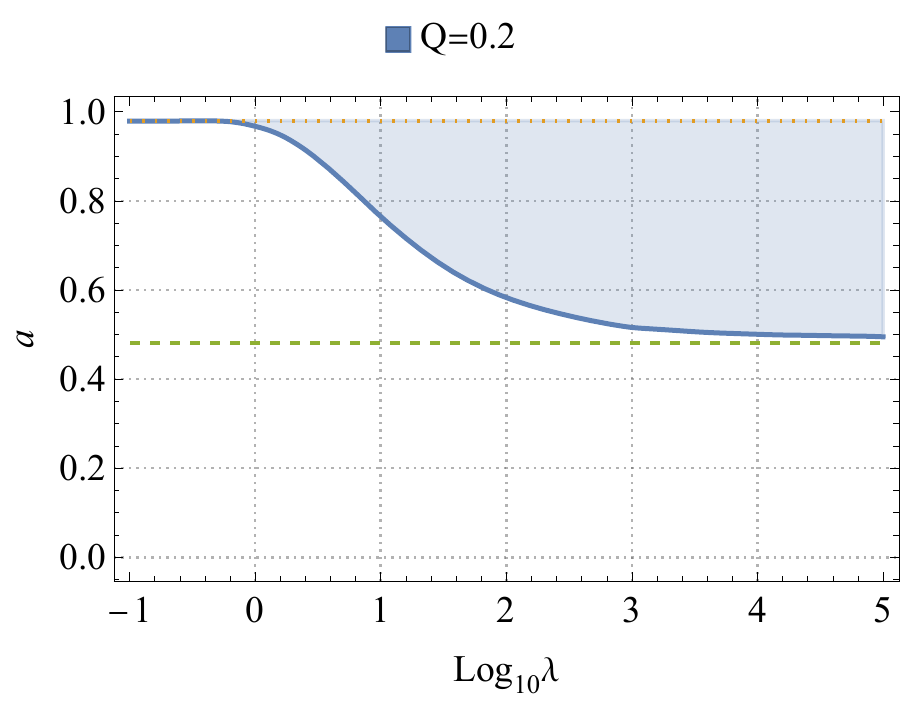}
\includegraphics[width=3in,angle=0]{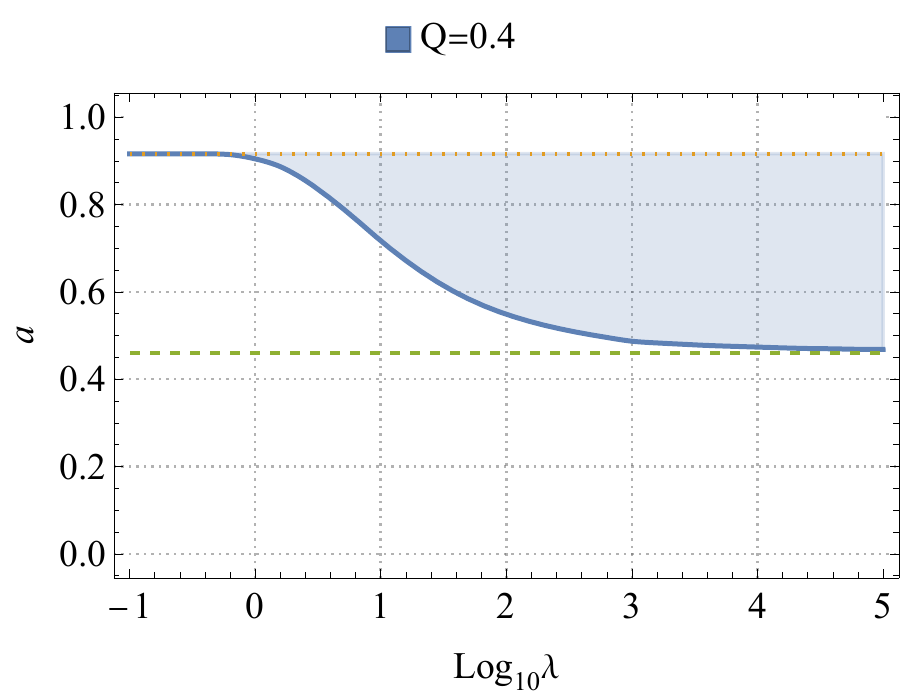}
\includegraphics[width=3in,angle=0]{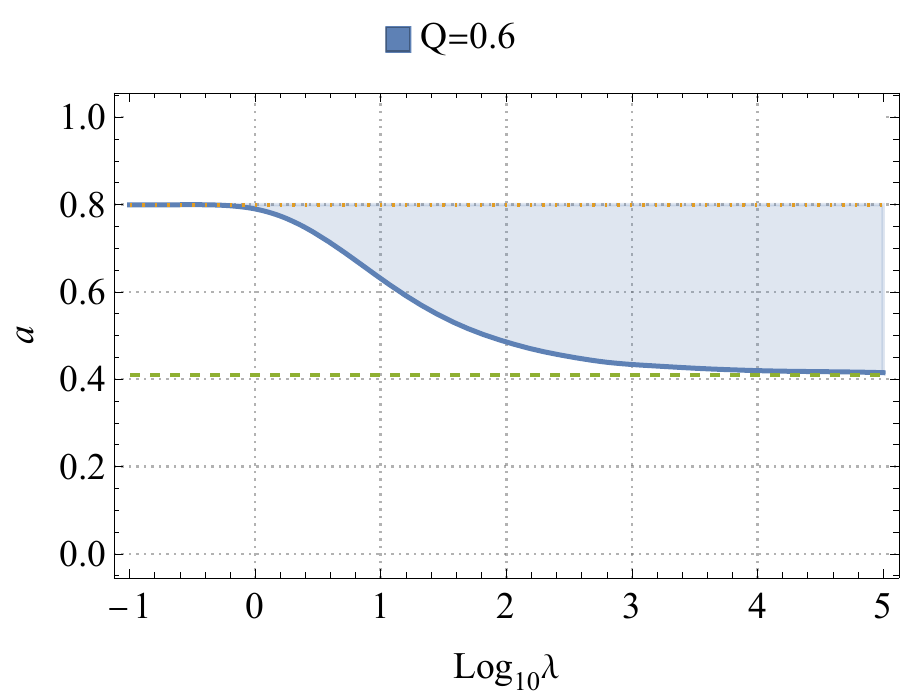}
\includegraphics[width=3in,angle=0]{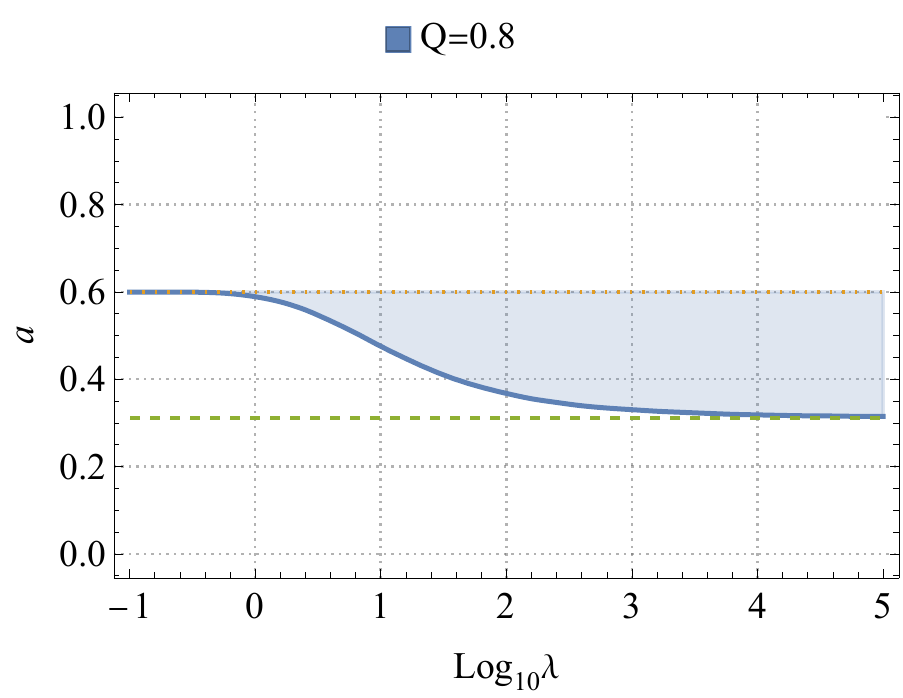}
\includegraphics[width=3in,angle=0]{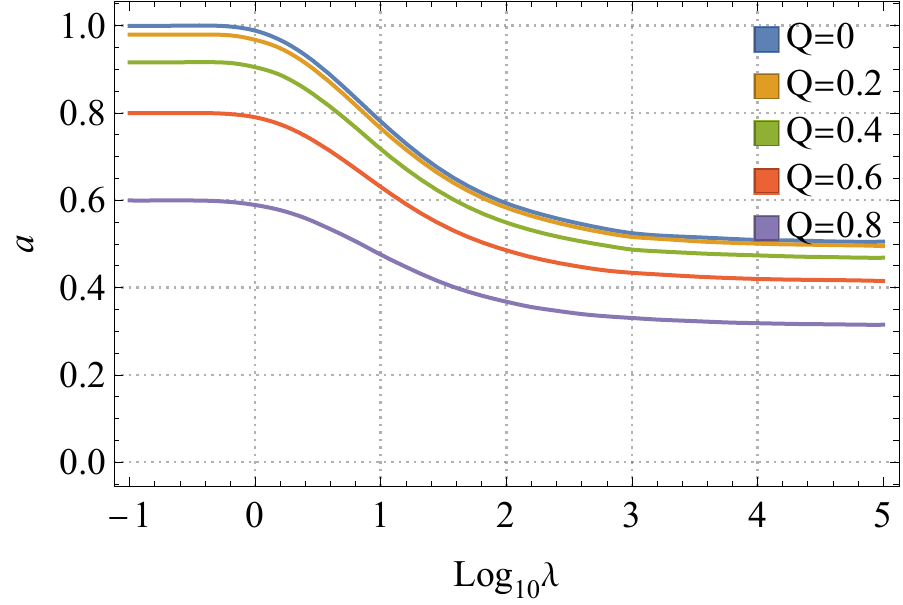}

\end{center}
\vspace{-5mm}
 \caption {The existence lines discriminate the parameter space spanned by $\lambda$ and $a$ with different $Q$ for $ M=1$ between the upper blue unstable region and the bottom stable region (between the horizontal dashed green line and the solid blue existence line). We have also plotted the lower and upper bounds of $a$. Note that $Q$ here in the diagram is in fact chosen critical value $Q_c$ in Fig. \ref{aqm}, and the value of $a$ corresponding to the horizontal dashed green line is in fact critical value $a_c$ in Fig. \ref{aqm}.}
 \label{teops}
\end{figure*}

Note that as we are dealing with scalar perturbations on  a rotating spacetime background, experience tells us that different $l-$led modes become coupled to each other, even with a specifically chosen initial $l$, as different $l-$modes with the same azimuthal mode  number $m$ will be excited. The $l=|m|$ mode will play a dominant role at the late times if we consider the stable modes, as pointed out in Ref. \cite{Doneva:2020nbb}. To be specific, we will consider spherical $l=m=0$ mode in what follows. In fact,  $m=0$ mode owns the shortest growth times and is  most relevant to the instabilities of the KN black hole. Among the dominant $m$ modes, the unstable region for the $m=0$ is the largest one. A similar situation is shown for the Kerr black hole \cite{Dima:2020yac,Doneva:2020nbb} in the Gauss-Bonnet gravity and for the Chern-Simons gravity \cite{Gao:2018acg}. Therefore, in the following, we only consider the field equation  (\ref{tpoeq}) with $m=0$ and the initial condition with $l=m=0$. In the FPR method, this initial condition becomes
\begin{equation}
\tilde{\psi}_i(t=0, x) \sim e^{\frac{-\left(x-x_c\right)^2}{2 \sigma^2}}.
\end{equation}

We show the time evolution of the dominant mode $l=m=0$ of the scalar perturbation wave function $\psi(t, x, \theta)$ at $x=0$ and $\theta=\pi/2$ with different coupling constant $\lambda$ in Fig. \ref{eld} by using two different numerical methods which ensure our results are correct. Here we set $x=0$ to  correspond to $r=r_0$ which is the maximal point of $\Delta/(r^2+a^2)$. By the time-domain behavior of the scalar perturbation, we get to know that the tachyonic instability, which is the hallmark of spontaneous scalarization of the KN black hole in the linearized regime, depends on $\bar{\lambda}, \bar{a}, \bar{Q}$ [see Eq. (\ref{redu})].

We also show the critical lines dividing parameter space into unstable regions which correspond to scalarized  black holes and stable regions, which correspond to bald KN black holes in Fig. \ref{teops}, which are dependent on the characteristic parameters $\bar{a},\,\bar{Q}\,,\bar{\lambda}$ of black hole-field system. In our simulations, we reach up to $\bar{\lambda}\sim 10^5$ and find the numerical results of the lower bound for critical $\bar{a}$  are in good agreement with the analytical results obtained from Eq. (\ref{bblef}). The upper bound of $\bar{a}$ corresponds to the extremal KN black hole with $\bar{a}^2+\bar{Q}^2=1$.

On the other hand, from Fig. \ref{teops}, we see that near large coupling regime $\lambda\gg 1$, the critical values of $a$ (or $Q$) decrease with increasing $\lambda$ for fixed $Q$ (or $a$). This qualitative behavior is completely compatible with the analytical result about the existence line in Eq.  (\ref{exlal}).

\section{Concluding remarks}\label{sec:con}
We have shown that a negative coupling between a scalar field and the GB invariant can lead to tachyonic  instability for  the KN black hole, which is the hallmark of spontaneous scalarization of the black hole in the linearized regime. Firstly, we analytically calculated critical black hole angular momentum and electric charge, which are the minimal necessary  values for which spin-charge scalarization occurs in the infinitely large coupling limit that corresponding to onset of the instability, see Fig. \ref{aqm}. Then we further corrected the result in the large but finite coupling regime to obtain the existence line Eq. (\ref{exlal}) which is the boundary of the bald KN black hole and the hairy scalarized black hole. It was revealed that for fixed  spin (charge), the characterized charge (spin) that triggers tachyonic  instability decreases with increasing coupling strength between the scalar field and the GB invariant. Finally, our analytical results were numerically verified and we carried out the whole threshold curve with the negative coupling parameter varying in its domain. Both of our analytical and numerical results are compatible with the ones in Refs. \cite{Doneva:2020nbb,Dima:2020yac,Hod:2020jjy} in the vanishing electric charge limit. An extra lesson is that, we found in the vanishing spin limit, the spontaneous scalarization of the  RN black hole can be triggered in the domain $\bar{Q}\gtrsim 0.957$.

\section*{Acknowledgements}
M. Z. is supported by the National Natural Science Foundation of China with Grant No. 12005080.  J. J.  is supported by the National Natural Science Founda- tion of China with Grant No. 12205014, the Guangdong Basic and Applied Research Foundation with Grant No. 217200003 and the Talents Introduction Foundation of Beijing Normal University with Grant No. 310432102.

%%%%%%%%%%%%%%%%%%%%%%%%%%%%%%%%%%%%%%%%%%%%%%%%%%%%%%%%%%%%%%%%
%%%%%%%%%%%%%%%%%%%%%%%%%%%%%%%%%%%%%%%%%%%%%%%%%%%%%%%%%%%%%%%%
\bibliographystyle{jhep}
\bibliography{refs}
%%%%%%%%%%%%%%%%%%%%%%%%%%%%%%%%%%%%%%%%%%%%%%%%%%%%%%%%%%%%%%%%
%%%%%%%%%%%%%%%%%%%%%%%%%%%%%%%%%%%%%%%%%%%%%%%%%%%%%%%%%%%%%%%%
\end{document}